\documentclass{ewic-latex}
\usepackage{natbib}
\usepackage{booktabs}
\usepackage{multirow} 
\usepackage{xurl}
\usepackage[hidelinks]{hyperref}
\renewcommand\harvardurl[1]{\textbf{URL:} \url{#1}}

\begin{document}

\runningheads{Peng $\bullet$ Russell-Rose}{EEG-based Word Association Paradigm for Adult ADHD Screening: An Exploratory Pilot Study}

\conference{Preprint}

\title{EEG-based Word Association Paradigm for Adult ADHD Screening:\\
An Exploratory Pilot Study}

\authorone{Caroline Peng\\
Goldsmiths, University of London\\
8 Lewisham Way\\
London, SE14 6NW\\
UK\\
\email{cpeng006@gold.ac.uk}}

\authortwo{Tony Russell-Rose\\
City St George's, University of London\\
Northampton Square\\
London, EC1V 0HB\\
UK\\
\email{Tony.Russell-Rose@citystgeorges.ac.uk}}

\begin{abstract}
With the prevalence of Attention Deficit Hyperactivity Disorder (ADHD) over the past decades, healthcare systems across the globe face critical diagnostic challenges due to long diagnostic waiting times and a reliance on subjective behavioural assessments that cannot distinguish ADHD from comorbid psychiatric disorders, especially for adult patients. This exploratory study investigates whether EEG-based word association paradigms show promise as a complementary approach to screening for ADHD in adults. Using a mixed-method approach, the study examines neurological and cognitive differences between neurotypical individuals (NT), clinically diagnosed ADHD participants (ND), and self-reported ADHD cases awaiting formal diagnosis (SR) across three word association tasks. While established EEG biomarkers (ERP N400, Theta/Beta Ratio, and Alpha Suppression) show no significant group differences, semantic distance analysis reveals a statistically significant main effect ($p=0.003$), with the SR group showing the most divergent associations. These preliminary findings suggest that word association paradigms may capture cognitive differences not detected by standard EEG metrics and encourage further large-scale investigation as a potential complement to existing adult ADHD screening tools.

\end{abstract}


\maketitle

\section{Introduction}
ADHD affects between 2.5\% and 6.7\% of the adult population globally and has been increasingly recognised as a lifelong neurodevelopmental disorder \citep{song2021prevalence}. Despite its prevalence, access to diagnosis remains severely constrained, with waiting times exceeding 12 months in many healthcare systems due to the long-standing view of ADHD as a childhood condition, leading to a shortage of certified clinical professionals specialised in adult ADHD \citep{Sibley2022,Sibley2018}.
~\\
The current screening tool, the ADHD Self-Report Scale (ASRS), is efficient but inherently subjective and struggles to differentiate ADHD from comorbid psychiatric disorders due to overlapping symptoms \citep{Caye2016,Ustun2017,Kooij2019,Lewczuk2024,Choi2022}. This situation is further complicated by social media, where the majority of content regarding ADHD self-assessment is misleading \citep{Verma2025}. There is an urgent need for objective screening tools that provide neurological validation for reported symptoms.
Recent studies have explored EEG-based biomarkers for ADHD diagnosis, but these are typically applied post-diagnosis in lab settings rather than during the initial screening stage. Several EEG studies have demonstrated that word association paradigms can reveal neurological differences between neurotypical and neurodivergent individuals when the tasks require creativity and originality \citep{White02072016,Planchuelo2022,Jauk2012}.
~\\
This exploratory study examines whether EEG-based word association tasks show potential as an alternative, complementary screening approach for adult ADHD by identifying neurological and semantic processing differences between neurotypical and neurodivergent individuals. Given the small sample size (N=15), we position this work as a pilot study intended to find promising directions and inform the design of future large-scale studies.

\section{Background and Related Work}
\subsection{Diagnostic Challenges}
The prevalence of ADHD has grown from 6.1\% to 10.2\% in the past decades, with 2.58\% persistent and 6.76\% late-onset among adult cases and is expected to continue increasing, yet the diagnostic infrastructure remains focused on childhood presentations \citep{song2021prevalence,Abdelnour2022,Sibley2022}. 
~\\
The current diagnostic framework for adult ADHD consists of two stages: initial screening followed by clinical assessment conducted by certified psychiatrists. However, the ASRS questionnaire lacks the ability to distinguish ADHD from other psychiatric disorders with overlapping symptoms, such as depression or anxiety \citep{Sobanski2007,Choi2022}. Furthermore, social media misinformation has also complicated public understanding of adult ADHD, leading to increased inappropriate self-reporting cases and referrals that are further straining healthcare services \citep{Verma2025}.

\subsection{EEG Biomarkers for ADHD}
EEG offers a cost-effective, non-invasive method for identifying objective neurological biomarkers \citep{Tripp2009,Chen2023}. One of the most studied and debated biomarkers is the Theta/Beta Ratio (TBR), with ADHD individuals typically showing elevated TBR compared to neurotypical controls \citep{Snyder2015,Ogrim2012}. Event-related potential (ERP) studies have also been suggested to be useful in identifying ADHD characteristics \citep{Hadas2021,Wiegand2016}. Alpha wave (8-12Hz) is also known to be modulated during attention and has also been considered as a candidate biomarker for ADHD, especially with its abnormal oscillations that have been repeatedly observed among those with ADHD \citep{Deiber2020}. 
~\\
Recent advances in EEG analysis techniques have further enhanced the potential clinical utility of neurological biomarkers. However, clinical adoption is hindered by high inter-subject variability and inconsistent replication across studies \citep{Müller2020}. Given the ADHD symptom trajectories fluctuate across adulthood, with only 9.1\% of childhood cases showing full recovery and 63.8\% experiencing periods of remission and recurrence \citep{Sibley2022}, adult populations may present greater neurobiological heterogeneity than the childhood samples on which most EEG biomarker research is based.

\subsection{Word Association and ADHD}
Word association paradigms have emerged as valuable tools for investigating cognitive and semantic processing differences, particularly in neurodivergent populations. White and Shah's study provided crucial evidence that individuals with ADHD showed different patterns of semantic processing during word association tasks \citep{White02072016}. Using both free association and controlled oral word association tasks (COWAT), they found that ADHD participants showed greater semantic distance in their word associations compared to neurotypical controls. This finding suggests that individuals with ADHD access more diverse and distantly related semantic concepts during associative thinking. 
~\\
\cite{Jauk2012} used EEG to investigate the neural correlates of creative word association in neurotypical populations. They found that creative word association tasks elicited specific patterns of alpha oscillations, particularly in frontal and temporal regions associated with executive control and semantic processing.
~\\
The convergence of behavioural and neural evidence suggests that word association tasks provide a promising opportunity for investigating ADHD-related cognitive differences.

\section{Methodology}
This study employs mixed methods combining quantitative EEG analysis and qualitative participant feedback to evaluate the potential of word association paradigms as a complementary approach to screening tools for adult ADHD. The primary investigation uses three distinct word association tasks conducted during EEG recording, complemented by post-study semi-structured interviews to gather participant insights about their experiences and perspectives. This approach examines both the neurological patterns and the user experience factors for assessing the feasibility of the word association test as a potential complement to existing screening approaches.

\subsection{Participants}
Fifteen adult participants were recruited from university networks: neurotypical individuals serving as controls (NT, $n=5$), clinically diagnosed ADHD individuals (ND, $n=5$), and self-reported ADHD individuals currently registered on the NHS waiting lists (SR, $n=5$).
~\\
To minimise linguistic bias, all participants had either native English proficiency or bilingual competency with English acquisition before the age of 6. Groups were balanced for gender to account for historical diagnostic disparities \citep{Fairman2020,Wernersson2020}. All participants completed the ASRS to verify alignment with their assigned groups. Medication status and psychiatric comorbidities were not systematically recorded or screened due to privacy concerns, which is acknowledged as a limitation.

\subsection{Procedure}
Participants were asked to perform three WATs during a 15-minute EEG recording, followed by a 10-minute semi-structured interview exploring attentional engagement, task experience, and views on the use of EEG.
~\\
The structured 15-minute EEG recording session was divided into three phases: pre-task resting state recording of 30 seconds with eyes closed followed by 30 seconds with eyes open to establish individual baseline measurements; three WATs with 30-second rest intervals between tasks; and a post-task resting state recording with eyes open and closed conditions lasting 30 seconds each.
~\\
Word cues and pairs for Task A (free association) and B (paired association) were systematically selected from the Small World of Words (SWOS) database and filtered by Shannon entropy values and Concreteness ratings to ensure the selection of commonly known words capable of eliciting diverse responses across participants \citep{Dedeyne2019,Brysbaert2014}. Task A and B each comprised 20 trials per participant, while Task C (COWAT) generated 30-80 word responses per participant depending on individual task performance. This design produced a substantial number of trials per participant for both EEG and semantic analysis. The three WATs were performed in randomised order to exclude priming effects and systematic bias \citep{Neely1991} and were implemented through a custom-designed website published on GitHub, ensuring standardised stimulus presentation and response collection across all participants.

\subsection{Data Analysis}
EEG data were recorded using ANT Neuro's eego software, with either 24-channel saline-based or 32-channel gel-based neurocaps depending on lab equipment availability. Both configurations adhered to the international 10/20 system electrode placement standards, and analysis focused on the 24 overlapping channels common to both setups. From these channels, biomarker-specific regions of interest were selected for analysis: the N400 and alpha suppression were analysed at the centro-parietal site Pz, where the N400 is maximal \citep{Klem1999}, while the Theta/Beta Ratio was analysed at the frontal sites Fz, F3, and F4, consistent with the frontal localisation of TBR and the frontal-lobe differences most reliably reported in ADHD \citep{Tripp2009}.
~\\
Data were sampled at 500 $Hz$, and an amplitude threshold of 70 $\mu V$, with artefact rejection and ICA applied for preprocessing. After artefact rejection, each participant retained at least 30 trials across three tasks, providing a sufficient trial count for reliable analysis at the individual level despite the small number of participants. 
~\\
Semantic distance for WATs was processed using GloVe word embeddings \citep{pennington-etal-2014-glove}. GloVe was selected for its established use in computational semantics research and its capacity to quantify associative distance in a standardised, reproducible manner. However, it should be noted that corpus-derived distributional semantics capture statistical co-occurrence patterns in language rather than individual-level semantic network structure and may miss non-linguistic aspects of meaning, such as imagery and emotional association, that free association tasks can elicit \citep{Kumar2021}. 
~\\
Within-group analysis examined individual variability patterns and response consistency across different word association tasks. Between-group statistical comparisons employed one-way ANOVA for each EEG metric, followed by post-hoc tests to identify specific group differences.
~\\
Given the exploratory aim of this pilot, analyses across the multiple EEG channels and tasks were not corrected for multiple comparisons; reported $p$-values should therefore be interpreted as indicative rather than confirmatory, and the effect sizes are emphasised accordingly.
~\\
Interview transcripts underwent systematic thematic analysis employing hybrid coding methodologies that combined predetermined categories (task difficulty, attention awareness, and assessment acceptability) with inductive identification of emergent themes.

\section{Results}
\subsection{EEG Results}
Valid EEG data from 14 participants were collected across three groups, with 4 from the NT group and 5 each from the other two groups. Contrary to the hypotheses, none of the EEG biomarkers reached statistical significance ($p>0.05$). While each participant contributed at least 30 clean trials, providing reliable individual-level EEG analysis, the between-group comparisons were conducted at the participant level, limiting statistical power to detect moderate group differences. 
~\\
ERP N400 analysis from Pz and Fz channels showed no significant group differences and no significant main effect of group, despite a large effect size ($\eta^2 = 0.162$), with the ND group exhibiting the highest amplitude and the SR group the lowest.
~\\
The TBR results from Fz, F3, and F4 channels contradicted established ADHD biomarker literature, which typically reports elevated TBR in diagnosed ADHD populations compared to controls, with the highest mean TBR in the SR group, followed by ND and NT. Analysis revealed no significant group differences and no significant main effect of group, with only the F4 channel showing a larger effect size of $\eta^2 = 0.240$.
~\\
Alpha suppression analysis from the Pz channel during word association tasks compared with resting states showed distinct patterns across groups, though not reaching statistical significance, with the ND group showing the strongest suppression, while the SR group showed the weakest.

\begin{table*}[ht]
  \centering
  \caption{EEG biomarker group statistics and one-way ANOVA results across channels (ERP N400, Theta/Beta Ratio, and Alpha Suppression). No metric reached significance at $p<0.05$.}
  \label{tab:eeg_results}
  \small
  \begin{tabular}{llrrrrrr}
    \toprule
    Biomarker & Channel & Group & Mean & SD & SEM & $F$ / $p$ & $\eta^2$ \\
    \midrule
    \multirow{6}{*}{ERP N400}
      & \multirow{3}{*}{Pz} & ND & -2.775 & 2.066 & 0.924 & \multirow{3}{*}{$1.066$ / $.378$} & \multirow{3}{*}{0.162} \\
      &                     & NT & -1.996 & 0.755 & 0.378 & & \\
      &                     & SR & -1.182 & 1.868 & 0.835 & & \\
    \cmidrule(l){2-8}
      & \multirow{3}{*}{Fz} & ND & 1.911 & 2.571 & 1.150 & \multirow{3}{*}{$0.017$ / $.984$} & \multirow{3}{*}{0.003} \\
      &                     & NT & 1.653 & 3.764 & 1.882 & & \\
      &                     & SR & 1.614 & 1.953 & 0.873 & & \\
    \midrule
    \multirow{9}{*}{TBR}
      & \multirow{3}{*}{Fz} & ND & 5.162 & 3.697 & 1.653 & \multirow{3}{*}{$0.155$ / $.858$} & \multirow{3}{*}{0.027} \\
      &                     & NT & 4.746 & 1.876 & 0.938 & & \\
      &                     & SR & 5.800 & 2.528 & 1.131 & & \\
    \cmidrule(l){2-8}
      & \multirow{3}{*}{F3} & ND & 3.830 & 1.775 & 0.794 & \multirow{3}{*}{$0.225$ / $.802$} & \multirow{3}{*}{0.039} \\
      &                     & NT & 4.683 & 3.103 & 1.552 & & \\
      &                     & SR & 4.540 & 1.299 & 0.581 & & \\
    \cmidrule(l){2-8}
      & \multirow{3}{*}{F4} & ND & 4.239 & 1.246 & 0.557 & \multirow{3}{*}{$1.739$ / $.221$} & \multirow{3}{*}{0.240} \\
      &                     & NT & 4.178 & 1.181 & 0.590 & & \\
      &                     & SR & 5.837 & 2.031 & 0.908 & & \\
    \midrule
    \multirow{3}{*}{Alpha Supp.}
      & \multirow{3}{*}{Pz} & ND & -2161.6 & 4223.6 & 1888.9 & \multirow{3}{*}{$0.164$ / $.851$} & \multirow{3}{*}{0.029} \\
      &                     & NT & -1733.4 & 2659.5 & 1329.8 & & \\
      &                     & SR & -1071.0 & 1446.6 & 646.94 & & \\
    \bottomrule
  \end{tabular}
\end{table*}

\subsection{Semantic Distance}
\subsubsection{Task A: Free Association}
The free association task revealed significant group differences in semantic distance patterns ($F(2,12)=9.77$, $p=0.003$, $\eta^2 = 0.62$), with the most pronounced differences emerging between the ND and SR groups. Tukey HSD post-hoc tests confirmed that the SR group ($M=0.289$) demonstrated significantly greater semantic distance than the ND group ($M=0.241$, $p=0.002$, $d=-3.55$).
This pattern contrasts with previous research by \cite{White02072016}, who found that individuals with diagnosed ADHD produced associations with greater semantic distance compared to neurotypical controls. 

\subsubsection{Task B: Paired Association}
Analysis of semantic relatedness judgements showed different patterns compared to the free association task. One-way ANOVA revealed no significant group differences ($p = 0.915$, $\eta^2 = 0.015$), indicating a small and non-significant effect. All effect sizes between groups were negligible to small.

\subsubsection{Task C: COWAT}
The COWAT analysis examined sequential semantic distance at the word level, treating each generated word as a trial, with participants generating between 32 and 84 words each across three letter prompts. While the ANOVA did not reach statistical significance ($p = 0.234$, $\eta^2 = 0.067$), and the effect size was in the medium range, pair-wise effect sizes revealed meaningful differences, with the NT group demonstrating the strongest clustering with the lowest sequential distance. Individual performance varied substantially within the ND group, ranging from 43 to 84 total words, highlighting the within-group heterogeneity. 

\begin{table*}[ht]
  \centering
  \caption{Semantic distance group statistics and one-way ANOVA results across the three word association tasks. For Tasks B and C, lower distance indicates more clustering.}
  \label{tab:wat_results}
  \small
  \begin{tabular}{llrrrrr}
    \toprule
    Task & Group & Mean ($M$) & SD & $n$ & $F$ / $p$ & $\eta^2$ \\
    \midrule
    \multirow{3}{*}{A: Free Assoc.}
      & NT & 0.2669 & 0.0232 & 5 & \multirow{3}{*}{$9.773$ / $.003$} & \multirow{3}{*}{0.620} \\
      & ND & 0.2405 & 0.0152 & 5 & & \\
      & SR & 0.2894 & 0.0121 & 5 & & \\
    \cmidrule(l){2-7}
    \multirow{3}{*}{B: Paired Assoc.}
      & NT & 6.78 & 1.50 & 5 & \multirow{3}{*}{$0.089$ / $.915$} & \multirow{3}{*}{0.015} \\
      & ND & 6.91 & 0.56 & 5 & & \\
      & SR & 7.05 & 0.71 & 5 & & \\
    \cmidrule(l){2-7}
    \multirow{3}{*}{C: COWAT}
      & NT & 13.80 & 4.44 & 5 & \multirow{3}{*}{$1.506$ / $.234$} & \multirow{3}{*}{0.067} \\
      & ND & 16.33 & 4.13 & 5 & & \\
      & SR & 14.20 & 4.31 & 5 & & \\
    \bottomrule
  \end{tabular}
\end{table*}

\subsection{Post-study Interviews}
Semi-structured interviews revealed that all of the ND and SR participants reported identical attention difficulties during tasks, especially COWAT, while NT participants maintained consistent focus. 
80\% of ND participants viewed the WAT as a more objective screening tool compared to questionnaires. SR participants were split, with 40\% found it interesting but not reliable, while the rest expressed no clear preference. The NT group was most sceptical, with 80\% participants viewing the approach as too casual for clinical use, though one participant found it acceptable.

\section{Discussion}
\subsection{EEG Biomarkers}
Contrary to existing literature, established EEG biomarkers did not show statistical significance in differentiating groups. This result is consistent with the high inter-subject variability widely reported in adult ADHD EEG research, particularly within the ND group \citep{Müller2020}. The large effect size found in N400 ($\eta^2 = 0.162$) and TBR ($\eta^2 = 0.240$) suggests that while neurobiological differences compared to the control group may exist, these effects were not detectable at the current sample size.

\subsection{Semantic Distance}
The standout finding was that the SR group showed significantly higher ($p=0.003$) semantic distance than the ND group. Several explanations could be found for this pattern. One possibility is that self-reported ADHD individuals may exhibit the enhanced divergent thinking typically associated with ADHD more prominently than clinically diagnosed individuals, who may have developed compensatory strategies through treatment or therapeutic intervention. However, alternative explanations cannot be ruled out. The SR group may represent a self-selected population with higher symptom severity or stronger identification with ADHD traits, and some SR participants may have different underlying psychiatric disorders than the ND group. 
~\\
The divergence between EEG and semantic findings may reflect the different measurement scales of these approaches: EEG biomarkers capture real-time neural oscillations that are highly sensitive to moment-to-moment variability and recording artefacts, whereas semantic distance measures aggregate across multiple responses within a task, producing a more stable behavioural signal. This distinction may explain why a small-sample pilot test could detect semantic but not neurological differences because the behavioural measure was simply more robust to noise at this scale.

\subsection{Limitations}
The study’s primary limitation is the small sample size ($N=5$ per group), which limits statistical power and generalisability at the between-group level. However, it should be noted that the individual-level data are more substantial, with at least 30 clean EEG trials per participant after artefact rejection. The limitation lies in the number of participants contributing to group-level comparisons. 
~\\
Medications and comorbid psychiatric conditions were not recorded due to private concerns. Given that 80\% of adults with ADHD have at least one comorbid psychiatric disorder, unrecognised co-occurring conditions may have influence on results in both the ND and SR groups \citep{Sobanski2007}. 
~\\
Equipment variation between 24- and 32-channel neurocaps introduced an additional source of measurement inconsistency, though analysis was restricted to the 24 overlapping channels common to both configurations. 
~\\
Additionally, it should also be noted that GloVe-based semantic distance captures distributional patterns derived from large text corpora, which may not fully correspond to the structure of individual semantic networks. 
~\\
Future research should employ larger, more diverse samples with consistent methodological approaches and tighter clinical controls to validate these preliminary findings.

\section{Conclusion}
This pilot study examined whether EEG-based word association paradigms could show promise as complementary approaches to adult ADHD screening, addressing the critical healthcare challenge of prolonged diagnostic waiting times. While the results do not support the development of a simple EEG biomarker-based ADHD screening tool, they provide valuable preliminary insights into the complexity of ADHD characteristics and the limitations of current approaches to objective assessment.
~\\
The most significant finding was that word association tasks, particularly the free association task, revealed meaningful cognitive differences between groups that were not detected through traditional EEG biomarkers. This suggests that computational semantic measures may offer a more detectable signal than neural biomarkers at this sample size, though whether this advantage reflects greater robustness or scalability remains to be confirmed with larger samples. And the failure of established EEG biomarkers to differentiate between groups, despite adequate effect sizes in some analyses, highlights the substantial heterogeneity within ADHD populations.
~\\
The path toward objective ADHD screening remains complex, but this pilot research provides valuable initial insights into the cognitive and neurological diversity within ADHD populations and highlights the importance of considering individual differences in the development of assessment approaches. 
~\\
Future efforts toward improving ADHD diagnosis must account for this complexity while maintaining focus on the ultimate goal of reducing barriers to appropriate care for those who need it.

\bibliographystyle{agsm}
\bibliography{sample-base}

\end{document}